\documentclass[sigconf]{acmart}
\usepackage{listings}

\AtBeginDocument{%
  \providecommand\BibTeX{{%
    \normalfont B\kern-0.5em{\scshape i\kern-0.25em b}\kern-0.8em\TeX}}}

\copyrightyear{2024}
\acmYear{2024}
\setcopyright{rightsretained}
\acmConference[SIGCSE 2024]{Proceedings of the 55th ACM Technical Symposium on Computer Science Education V. 1}{March 20--23, 2024}{Portland, OR, USA}
\acmBooktitle{Proceedings of the 55th ACM Technical Symposium on Computer Science Education V. 1 (SIGCSE 2024), March 20--23, 2024, Portland, OR, USA}
\acmDOI{10.1145/3626252.3630871}
\acmISBN{979-8-4007-0423-9/24/03}

\begin{document}

\title{How We Manage an Army of Teaching Assistants: Experience Report on Scaling a CS1 Course}

\author{Ildar Akhmetov}
\authornote{Work performed while affiliated with the University of Alberta}
\email{i.akhmetov@northeastern.edu}
\orcid{0000-0002-6660-8890}
\affiliation{%
  \institution{Northeastern University}
  \city{Vancouver}
  \state{BC}
  \country{Canada}
}

\author{Sadaf Ahmed}
\email{sadaf@ualberta.ca}
\orcid{0000-0001-6540-3909}
\affiliation{%
  \institution{University of Alberta}
  \city{Edmonton}
  \state{AB}
  \country{Canada}
}

\author{Kezziah Ayuno}
\email{ayuno@ualberta.ca}
\orcid{0000-0002-4722-5210}
\affiliation{%
  \institution{University of Alberta}
  \city{Edmonton}
  \state{AB}
  \country{Canada}
}

\renewcommand{\shortauthors}{Akhmetov, Ahmed, and Ayuno}

\begin{abstract}
A considerable increase in enrollment numbers poses major challenges in course management, such as fragmented information sharing, inefficient meetings, and poor understanding of course activities among a large team of teaching assistants. To address these challenges, we restructured the course, drawing inspiration from successful management and educational practices. We developed an organized, three-tier structure for teams, each led by an experienced Lead TA. We also formed five functional teams, each focusing on a specific area of responsibility: communication, content, "lost student" support, plagiarism, and scheduling. In addition, we updated our recruitment method for undergraduate TAs, following a model similar to the one used in the software industry, while also deciding to mentor Lead TAs in place of traditional training. Our experiences, lessons learned, and future plans for enhancement have been detailed in this experience report. We emphasize the value of using management techniques in dealing with large-scale course handling and invite cooperation to improve the implementation of these strategies, inviting other institutions to consider and adapt this approach, tailoring it to their specific needs.

\end{abstract}





\maketitle

\section{Introduction}

Over the past few years, we have witnessed rapid growth in the enrollment numbers for our CS1 course (CMPUT 174) at the University of Alberta. As the course expanded, we faced a major challenge in managing a horizontally structured team of more than 40 teaching assistants (TAs). This large scale led to fragmented and anecdotal information sharing, awkward and inefficient weekly meetings, and poor understanding of lab activities. We sought inspiration from best practices in management and education to restructure the course and achieve the following goals:

\begin{enumerate}
    \item \textit{Effectively scale the course} to accommodate potential enrollment growth of several thousand students without compromising the quality of students’ experience.
    \item Establish efficient \textit{task delegation} from instructors to TAs while ensuring high-quality work.
    \item Promote \textit{leadership development for TAs} and encourage their initiative in driving long-term course improvements.
\end{enumerate}

To achieve these goals, we developed a three-tier organizational structure where each TA is part of a small team led by an experienced Lead TA. Instructors, in turn, oversee Lead TAs, promoting efficient information flow. We created two team categories: \textit{regular teams}, which only take care of regular (routine) tasks such as office hours, grading and oral code walks, and \textit{functional teams}, which handle specific administrative duties in addition to the above mentioned regular tasks. The five functional teams were: Communication Support, Content Support, "Lost Student" Support, Plagiarism Support, and Scheduling Support. Each team implemented innovative practices, which we discuss in detail in this report. We also touch on the implementation of an undergraduate TA recruitment method that is akin to that used in the software engineering industry \cite{spolsky2007smart}, and the "train the trainer" approach we used instead of establishing a comprehensive TA training program.

The poster \cite{akhmetov2022managing} we presented at SIGCSE 2023 initiated insightful conversations, emphasizing the importance of a detailed experience report. In this report, we carefully examine the course management approach utilized during Fall 2022 and share our reflections on the outcomes. We also briefly discuss our unsuccessful attempt to replicate this course management design for a CS2 course with a similar enrollment size in Winter 2023. The urgency of carrying out the course and ensuring the practical implementation of these decisions limited thorough formal research efforts (except for \cite{mcdonald2023}, where we analyzed TAs’ experiences in the course). As a result, our lessons learned mainly stem from our own first-hand experiences (all three authors worked on the course redesign, with the first two authors serving as instructors and the third author serving as an undergraduate TA) and author-led retrospectives with the TAs. We leave the planning and execution of several formal studies for future work.

\section{Related Work}

Most of the existing research on handling large CS classes targets the issues presented by growing class sizes, favoring pedagogical rather than managerial perspectives. The methods that researchers have looked at include peer learning  \cite{porter2011peer}, creative seating plans in classrooms \cite{minnes2018lightweight}, smaller learning units called "micro-classes" \cite{alvarado2017micro}. Other successful strategies include using personal tutors or mentors and different ways to organize group work \cite{minnes2018lightweight, pieterse2014managing}.

Attempts to successfully scale introductory CS courses highlight the importance of the role of teaching assistants in the success of scaling efforts \cite{forbes2017scaling, mirza2019undergraduate, mohandas2020effectiveness}, which fully aligns with our approach. This underscores the need for effective TA recruitment and training, and this aspect of our work draws inspiration from studies such as \cite{komarraju2008social, smith2021developing, dechenne2015modeling, lane2021motivating, riese2022training, estrada2017bridging}. Another successful scaling strategy involves automating some aspects of course management \cite{wilcox2015role, skalka2020automated, macwilliam2012scaling} and utilizing learning analytics \cite{hui2017can, azcona2018personalizing, quille2022press}. These approaches have encouraged our functional teams to consistently invest in automating their processes and opt to make more data-driven decisions.

Our organizational structure was inspired by the UC Berkeley Data 8 team structure \cite{data8_pedagogy_guide}, whose method of structuring teams and tiered TA roles (AI/Tutor/uGSI) served as a starting point for our approach. Each functional team within our structure draws inspiration from the existing research in their fields. For instance, support strategies for "lost" students are explored in \cite{gordon2023ultra, ott2016translating, petersen2016revisiting}; effective course communication has been investigated by \cite{bridson2022delivering, vahid2020online, johnson2022embracing}; and plagiarism detection caught our attention primarily in the context of detection tool development \cite{ahadi2019comparison}.

Finally, our work would not be possible without incorporating management theories into the educational context. Essential resources such as \cite{fitzpatrick2015debugging, mcgrath2017little, fournier2017manager, hoekman2016new, thomas2019pragmatic, demarco2013peopleware, stanier2020become, rooke2005seven} not only informed our course design but also played a crucial role in effectively mentoring Lead TAs and fostering their growth as leaders.

\section{Course Management}

\subsection{Organizational Structure}

\begin{figure*}[h]
  \centering
  \includegraphics[width=\textwidth]{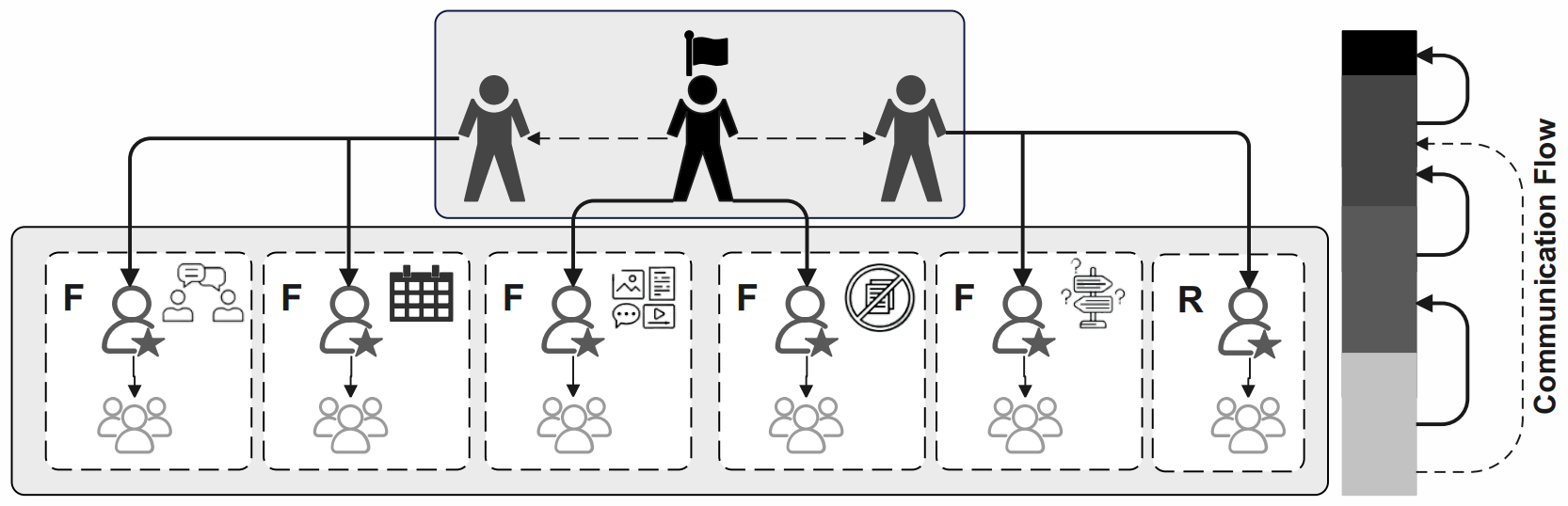}
  \caption{Course Organizational Structure}
  \label{fig:orgstructure}
\end{figure*}

Following the "two-pizza rule"\footnote{The "two-pizza rule" is a guideline that suggests a team should be small enough to be fed by two pizzas, meaning it promotes smaller, more efficient teams.} \cite{denning2019amazon}, we crafted a scalable three-tier organizational structure, shown in Figure~\ref{fig:orgstructure}. The top tier is made up of course instructors, one of whom assumes the special role of Course Coordinator. Each instructor oversees several Lead TAs (experienced graduate TAs with strong leadership potential) who, in turn, manage small teams of 5 to 6 members each. Some teams are \textit{functional} (marked \textbf{F}), while the rest are \textit{regular} (marked \textbf{R}). Although the illustration shows only one regular team, in practice, we had three such teams, and the structure can be efficiently scaled by adding as many regular teams as necessary based on the number of TAs.

The communication flow was designed to foster shorter, more focused meetings, promoting efficient information exchange. On Mondays, each team held separate meetings at convenient times for all members. These meetings, led by Lead TAs, covered status updates for office hours, labs, and functional tasks and included some training activities. On Tuesdays, instructors met with the Lead TAs they supervised to discuss issues reported by team members, often incorporating elements of brainstorming and leadership coaching. Finally, all instructors convened on Wednesdays to share information gleaned from the TAs and make high-level course decisions. While TAs were encouraged to contact their Lead TAs first, a dedicated Discord channel was available to expedite urgent inquiries (shown as a dashed communication line on the chart).

\subsection{TA Time Allocation}

Introducing functional tasks for TAs required a reassessment of how their weekly time commitments were allocated. With two types of teams (functional and regular), two levels of TAs (Lead TAs and team members), and three types of contracts based on weekly time commitment (6 hr, 9 hr, and 12 hr), we could potentially define twelve distinct TA time commitment profiles. These profiles served as a foundation, with adjustments made to each TA’s actual hours to balance the demands of the course and individual functional teams. Table~\ref{tab:time_profiles} showcases the time commitment profiles for 6 hr and 12 hr contracts.

\begin{table}
  \caption{TA Time Allocation Profiles (hours per week)}
  \label{tab:time_profiles}
  \begin{tabular}{lcccccc}
    \toprule
   Profile &Regular  &Meetings    &Func. &Training  \\
           &tasks    &(attending /&tasks      &(participating /\\
           &         &leading)    &           &facilitating) \\
    \midrule
    FuncLead12      &4    &1 / 1      &4      &0 / 2\\
    FuncMember12    &8    &1 / 0      &2.5    &0.5 / 0\\
    FuncMember6     &2    &1 / 0      &2.5    &0.5 / 0\\
    RegLead12         &9    &0.5 / 0.5  &0      &0 / 2\\    
    RegMember12       &11   &0.5 / 0    &0      &0.5 / 0\\
    RegMember6        &5    &0.5 / 0    &0      &0.5 / 0\\
    \bottomrule
  \end{tabular}
\end{table}

\subsection{Functional Teams}

\subsubsection{Communication Support}

The students’ learning experience greatly depends on the support they obtain from the instructional team; however, in a large course, it can quickly become difficult to manage the volume of students’ questions. In our past experience, we assigned all TAs to manage discussion forums, but in reality, only some participated, and their contributions were noticeably inconsistent. To solve this, we created the Communication Support team, which was responsible for handling students’ questions on the official course Discord server. Additionally, the team oversaw and maintained the server, setting up channels, roles, and permissions. During the initial weeks, the team aimed to cultivate a positive learning environment that encouraged student collaboration and peer support by setting the tone and mood of the Discord community.

\subsubsection{Content Support}

Reviewing course content (lecture slides, lab assignments, quizzes, exams, etc.) is a standard responsibility for TAs. However, we previously observed that the most reliable and trusted TAs frequently took on such "extra" work, often without being relieved of their regular duties. Therefore, the primary reason for having this team was to establish a dedicated group for content review tasks. In addition to core reviewing responsibilities, the team handled various housekeeping duties related to labs, quizzes, and exams, such as downloading lab submissions from the learning management system (LMS), separating them by section, and uploading grades to the LMS.

\subsubsection{"Lost Student" Support}

This team was tasked with identifying students who might be struggling with the course and offering proactive support. To achieve this, we utilized various sources of data to identify lost students. First, we assumed that students joining the course late often feel lost. Therefore, during the initial two weeks, before the add/drop deadline, we aimed to identify students who: (a) joined the course late; (b) never accessed the LMS; or (c) never joined the official Discord server. After the add/drop deadline, on a weekly basis, the team used two methods to identify lost students: (a) \textit{proactive identification} using the LMS extraction script (developed by the team); and (b) \textit{reactive identification} through the Lost Student report form, which other TAs and instructors could use to report students who were struggling with the course.

The proactive LMS extraction script for lost student identification is written in Python using the Pandas framework. It parses the LMS data (we use a Moodle-based LMS) and selects students who either did not submit assignments or had grades below the cut-off for \textit{N} previous lab assignments, \textit{M} previous quizzes, or \textit{K} previous midterm exams (where \textit{N}, \textit{M}, and \textit{K}, as well as the cut-offs for different deliverables, can be adjusted). The script reads data from a CSV file, so it can be potentially used with any LMS.

Once a lost student was identified, they entered the pipeline. First, the case was processed by the "Lost Student" support team Lead TA. If the student had already been recently contacted, they were skipped for the time being. Otherwise, the case was assigned to a member of the team. The assigned TA then sent a friendly, casual email to the student, conveying empathy and offering assistance (the email templates were authored by the team and can be reused). If the student agreed to a meeting, they had a 15-30 min meeting (preferably in-person, on campus), and the TA reported the results to the Lead TA, who, in turn, presented aggregated results and selected cases to the supervising instructor.

\subsubsection{Plagiarism Support}

Regrettably, plagiarism is a common issue in CS1 courses. In the past, instructors had to manage the time-consuming tasks related to plagiarism detection. Therefore, the primary objective of this team was to alleviate some of that burden. Instead of using Moss, which we utilized previously, we successfully adopted compare50\footnote{https://github.com/cs50/compare50}, a robust open-source code plagiarism detection tool developed by the CS50 team at Harvard. The Plagiarism Support team ran compare50 the morning following each lab assignment deadline and analyzed its reports. Since similarity checking tools cannot consistently provide accurate numerical scores, the team delved deeper into the specifics of each lab assignment (e.g., understanding the amount of boilerplate code provided to students and identifying which lecture examples were acceptable to use) and prepared a report for instructors. After the instructors assessed the flagged cases, they could pursue some of them further. Along with weekly checking and reporting, the team maintained a spreadsheet to track the progress of all plagiarism cases, allowing instructors to easily access that information.

\subsubsection{Scheduling Support}

Scaling a course introduces significant scheduling challenges, which is familiar to business managers, but has previously been less recognized by educators. With 1,055 students, 45 TAs, 7 lecture and 41 lab sections, we encountered a considerable scheduling challenge. Additionally, we provided office hours every weekday from 8 am to 8 pm, both in-person and online. Since most hours were staffed by multiple TAs, we offered a total of 150 office hours weekly. To generate the initial schedule, we used a linear solver that produced a draft based on our constraints (primarily, lab schedules, the number of TAs required for each office hour and lab, and TA availability). The linear solver model is a crucial component of scheduling success; however, due to the scope of this paper, we will present it separately.

Although creating the initial schedule was a significant challenge, the process didn’t end there. TAs, like other employees, may get sick or encounter time conflicts. With over 200 TA "shifts" per week (office hours and labs combined), addressing the continuous stream of rescheduling requests and adjustments became crucial. This is where the Scheduling Support team truly excelled. If a TA couldn’t attend their scheduled office hour or lab, they contacted the team via a dedicated Discord channel. A team member would self-assign to the case, confirm the duration of the change, and encourage the TA to coordinate shift swaps with peers. If that wasn’t possible, the Scheduling Support TA actively searched for a replacement through Discord and personal emails. Once a replacement was found, they updated the records and set reminders to revert the schedule when necessary. The team was advised to use their best judgment to evaluate the reasonableness of requests while maintaining an efficient schedule. Luckily, many issues could be resolved by simply changing the modality from in-person to online.

Another crucial aspect of this team’s work, making it the real "HR department" of the course, involved monitoring and controlling TA attendance and workload. First, all cases of TA absences, reported by students or peers, were tracked by the team and reported to the instructors. Second, TAs were asked to report office hours with too many or too few students to the team so that schedule adjustments could be made. Lastly, we started the implementation of automatic control mechanisms. Specifically, we developed a script to analyze Zoom logs, enabling the team to flag TAs who were absent during their office hours, arrived late, or left early.

\subsection{TA Management}

\subsubsection{Recruitment and Onboarding}

Unfortunately, there is a prevailing assumption that any graduate student can TA an introductory course, so typically, those graduate students who are not requested by advanced courses get assigned to CS1/CS2. As a result, although we could request some experienced TAs for Lead TA positions, we were unable to implement any other recruitment procedures for them. In contrast, undergraduate TAs underwent a recruitment process that involved advertising the positions through campus ad boards and informal Reddit and Discord communities. Applicants completed a form containing a LeetCode problem, an open-ended question regarding a critical skill a TA should possess, and an instructional video task to assess their presentation abilities. Following selection, the onboarding process included an initial meeting with icebreakers to encourage collaboration among TAs, as well as providing documents to clarify their roles.

\subsubsection{Training and Mentoring}

Initially, we explored the idea of creating a dedicated Pedagogy Support team to support TA professional development. However, we ultimately chose to have each Lead TA be responsible for guiding their team members to become better teaching assistants; instructors mentored and coached Lead TAs in handling teamwork issues to help them grow as leaders. To allow easy replication of the course management structure, we created comprehensive documentation, including various TA guides (onboarding guide, lab guide, office hour guide, etc.), a distinct collection of resources for Lead TAs (onboarding guide, training guide, weekly meeting guide, weekly readings on leadership and management), and extensive instructions for each functional team. These materials can serve as a foundation for adoption by other institutions, if needed.

In terms of TA training, we admit that we only scratched the surface, since we primarily used flashcards based on csteachingtips.org materials \cite{twarek2022tricky}. On a weekly basis, Lead TAs were asked to choose a few flashcards featuring potential tricky situations and employ them to encourage group discussions among their team members.

\subsection{Scalability}

The organizational structure is designed to be highly scalable, capable of accommodating a course with up to a few thousand students. With an average team size of 6 TAs, 5 functional teams account for 30 TAs, while the remaining TAs can form a virtually unlimited number of regular teams. Given a student-to-TA ratio of 1:25, as seen in our course, each additional 150 students would necessitate an extra regular team. When enrollment grows larger, some functional teams may need duplication (e.g., having multiple Communication Support teams operating in the same functional area as independent units). Alternatively, as the workload increases, functional team TAs could be relieved of more regular duties like office hours or grading, potentially resulting in a 100\% functional workload.

\section{Lessons Learned and Future Directions}

\subsection{Functional Teams}

\subsubsection{Communication Support}

We experienced success with this functional area; students praised the positive energy, welcoming atmosphere, and learning-focused environment on the server. Numerous lessons were learned regarding Discord communication, which could potentially form the basis for a separate experience report. It required several iterations to establish an effective mechanism: appointing a team of three "week experts" for each week-long shift. Prior to their shift, the "week experts" met with one of the instructors for an in-depth review of the upcoming lab assignment. Then, they took care of all incoming questions during that week. The main downside of this approach was a higher workload during the shift. Although it balanced out over the course of the semester, it could feel overwhelming at times, particularly due to the mental strain of being "on call" for a week.

While the "week expert" system aided in distributing work more evenly among TAs, we still observed varying levels of activity on Discord among them. This variation could be attributed to differences in TAs’ personalities and English comprehension levels, with more extroverted, people-driven TAs and those with a stronger command of English excelling in this role. A valuable change to consider for the future would be allowing TAs the opportunity to select the functional team in which they feel they would perform best, preferably before the start of the semester.

\subsubsection{Content Support}

Looking back on this team’s experience, we saw the benefits of having a separate group for handling content review tasks. First, we noticed that tasks were well-balanced among team members, with the Lead TA managing this successfully. Second, focusing on just one functional area allowed the team to improve the review process and constantly increase quality.

Although this team tackled tactical content improvement tasks effectively, in the future, we aim to leverage their expertise for strategic, long-term course improvement processes. To achieve this, we need to establish efficient feedback mechanisms. For instance, the Communication Support team could summarize the most frequently asked questions about specific lab assignments and share them with the Content Support team to suggest improvements. Another approach could involve analyzing TAs’ office hour notes to identify confusing or overly complex content elements. The team’s recommendations would contribute to mid-season course development efforts.

\subsubsection{"Lost Student" Support}

The progress of this team was perhaps the most insightful. Starting with a blank slate, they had only a four-page document containing basic guidelines. The team created a list of helpful resources to share with students, authored five email templates, and developed the data pipeline for selecting lost students. Within the first two weeks, they supported 96 students who joined late and 63 students who never accessed the LMS. In the remaining weeks, they contacted 280 students (approximately 14\% of all enrolled students) who might have struggled with the course at some point, resulting in 83 meetings with students (conversion rate 30\%). 

The main (and unexpected) lesson learned was that some underperforming students did not want to be helped. We observed that some students were selected and reported to the team throughout the semester but never responded to the emails sent by the team. Another issue consistently reported was that one-on-one meetings tended to transform into tutoring sessions. The team lacked the resources to provide tutoring support; thus, they learned that the best value can be provided by encouraging students to use existing support resources efficiently (e.g., attending office hours at specific less popular times or using office hours to ask questions not directly related to lab assignments). Based on their iterative experience, the team developed a six-step meeting scenario that begins with an icebreaker and includes discussing efficient lab work strategies, introducing problem decomposition, explaining available resources and problem-solving methods, addressing any additional questions, and finally, providing a comprehensive helpful resources document post-meeting.

During the retrospective, the team expressed a desire to determine whether they truly provided value to students. Despite anecdotal evidence of value, unfortunately, we did not adequately track one-on-one meetings, so we cannot determine whether such interventions improved students’ performance. We hope to implement a properly designed study to assess this next year. Another idea suggested by the team was to introduce weekly or biweekly check-ins instead of one-off meetings. This approach could help build a better rapport between a TA and a student. However, we recognize that regular check-ins can be time-consuming, so an alternative could be small group tutoring sessions.

\subsubsection{Plagiarism Support}

The team’s workflow quickly became efficient, needing only minor adjustments. Our main finding was that there weren’t enough tasks for all five team members, as the Lead TA could handle most duties spending only a few hours weekly. Based on this, we are considering two potential approaches for the future. One option is to assign the Plagiarism Support role to a single dedicated TA, instead of creating an entire team. Another possibility is to give the team more responsibilities related to helping instructors with the Code of Student Behaviour process. These tasks could include setting up meetings with students, acting as a witness and note-taker during meetings, assisting in writing reports, and compiling documents. However, we understand that dealing with such sensitive matters would require special training.

\subsubsection{Scheduling Support}

We are proud to report successful implementation of the "HR department" of a large CS1 course. This team saved perhaps hundreds of hours of instructors’ time by overseeing the schedule of labs and office hours and handling day-to-day rescheduling requests.

One of the early challenges of this team was an unbalanced workload, where the Lead TA and one more experienced TA self-assigned themselves to most requests, while the rest of the team had nothing to do. This was solved by making each team member responsible for a specific weekday. Another challenge was last-minute swap requests or cancellations, where the team did not have enough time to find a replacement. We realized that a TA often requests to cancel an in-person office hour but is willing to do it online, which was a solution for some cases. However, we need to study existing business practices to mitigate this risk.

Monitoring TA attendance was limited yet valuable. When a TA was found to be absent or late during their office hours, an instructor contacted them. We observed that such mild interventions were effective since we did not notice any repeated red flags. Surprisingly, several absences occurred because TAs were unaware of their schedules, so in the future, we plan to automatically generate calendar events for each TA based on the schedule. Another opportunity is automating data collection and analysis. We acknowledge the limitations of our current monitoring efforts, since we could only track online office hours by parsing Zoom logs, while in-person office hours and labs were not consistently monitored. We brainstormed possible ways to implement in-person monitoring, but these approaches tend to be overly invasive and time-consuming. Again, we hope to adapt existing business practices to make this possible.

The main factor hindering the adoption of our scheduling approach is reliance on a large and complex spreadsheet, which is coupled to some degree with course-specific data sources. To make our scheduling system adaptable by other institutions, we started development of an open-source scheduling system that will integrate the functionalities of the scattered tools and scripts we currently use and offer additional schedule optimization functionalities.

\subsection{TA Management and Training}

Assigning more functional tasks to teaching assistants highlights the need for effective recruitment. Like in business, finding and assigning the right individuals is challenging. Our selection of Lead TAs was based primarily on prior experience working with each TA, and in retrospect, we believe this approach was successful as all Lead TAs thrived in their roles. In future course iterations, we aim to identify the best TAs and, later, promote them, making the system self-sustaining.

While we carefully assessed candidates for each Lead TA position, other TAs were randomly assigned to teams. Team retrospectives revealed that TAs who excelled often had their strengths aligned with their functional tasks. This suggests that assessing TAs’ personal preferences before assigning them to teams could enhance team assignments.

The start and end of a course are usually the busiest times for instructors. This makes it hard to set aside time for effective onboarding and offboarding. Effective onboarding should be more than just a "kick-off" meeting, so it’s vital for TAs to get access to course management documentation ahead of time. Similarly, offboarding should include retrospectives with the teams and proper knowledge transfer. For our reflections, we used the FLAT technique  \cite{flat2021} (standing for "Future direction, Lessons learned, Accomplishments, and Thank you"). It served as an excellent framework, enabling the team to reflect on their achievements and offer insights for the course’s next iteration.

Our approach to TA management showed considerable success, yet room for improvement remains, particularly in the area of TA training. First, we identified the need for improved shadowing processes: a system where novice TAs could learn from their more experienced peers by observing their labs and office hours (shadowing), or via direct mentorship and observation \cite{peermentorship}. Though requiring substantial organization due to the high number of TAs, such a process could significantly enhance their teaching prowess \cite{groom2006building}. Second, although flashcards showed generally positive results in initial usage \cite{twarek2022tricky}, many Lead TAs ceased using them after a while, since they became repetitious. To combat this, we’re considering the development of an e-learning course for TA training that can be taken asynchronously during the early weeks of the semester. Lastly, weekly readings on leadership and management - sourced from various text resources \cite{fitzpatrick2015debugging, mcgrath2017little, fournier2017manager, hoekman2016new, thomas2019pragmatic, demarco2013peopleware, stanier2020become, rooke2005seven} - were consistently shared with Lead TAs. However, without proper follow-ups and debriefing on these materials, the learning potential of these resources might not have been fully realized.

\subsection{Scalability}

Overall, we can proudly report that our proposed course management approach was adept at handling a large CS1 course featuring 1,000 students. No significant constraints were encountered at this level of enrollment, indicating that this organizational structure could potentially accommodate even larger student populations. We recognize that complications might occur if the course were taught by more than seven or eight instructors, possibly necessitating an additional layer of management. Given the context of our university, where each section typically includes about 200 students and each instructor teaches an average of two sections, expanding the organizational structure might be required if enrolment were to increase to around 3,200 students.

Our approach was crafted to efficiently upscale, but interestingly, we found that downsizing might not be feasible. In particular, for the winter term of the same CS1 course — which is smaller in scale, having only 365 students across two sections and managed by 16 TAs — the instructor chose not to implement our proposed system. Instead, the more traditional "flat" structure was retained to reduce overhead. This suggests the need to create a more streamlined version of our course management structure that is specifically tailored to accommodate smaller-scale classes.

\subsection{Replication Attempt in CS2}

After seeing success with the new organization structure in the CS1 course, one of us tried to use the same structure for a CS2 course. While certain parts, like the functional teams, worked to some degree, the implementation didn’t work as a whole. TAs, who were part of both CS1 and CS2, noted that communication between instructors and TAs in CS2 wasn’t as good, which sometimes left students uninformed. Some TAs also mentioned that they felt less updated and involved in CS2 compared to CS1, particularly praising the "level of organization and professionalism" of their CS1 experience.

We believe the main reason for these issues was the insufficient buy-in of instructors to the new structure in CS2. Unlike in CS1, where all instructors supported the new system and excelled in their roles, in CS2, two out of the three instructors did not seem fully on board. Considering this, we see the need for easy-to-understand, engaging course materials that cater not just to TAs, but also to instructors. Such materials could help them understand the course structure better and, therefore, be more open to adopting it. However, we understand that getting this level of buy-in is never easy and may require significant effort.

\subsection{Future Directions}

The purpose of this experience report is to share and reflect on our novel approach to restructuring the organization of a large-scale CS1 course. However, we recognize that this report only touches the surface. We enthusiastically invite other institutions to implement this structure, and we’re more than willing to share our internal documentation, guidelines for instructors and TAs, data processing scripts, and other resources with any instructors interested in adopting this strategy. Given the crucial role of instructor engagement in the successful implementation of our proposed approach, we earnestly hope to generate synergies and potentially construct an “off-the-shelf,” easy-to-use course management framework.

In addition, we can identify several potential research avenues and warmly welcome collaboration on any of these. First, the findings and assumptions mentioned in this report require proper validation. Second, we’re keen on extending the promising research \cite{mcdonald2023} related to examining TAs’ experiences. Last, understanding the experiences of functional teams strikes us as worthy of further exploration. For instance, a study could be based on the extensive data gathered by the "Lost Student" Support Team, examining the effects of early interventions and illuminating various facets of underperforming students’ behaviour. The data accumulated by the Scheduling Support Team could also reveal insights regarding TAs’ dedication and commitment to teaching.

\section{Acknowledgments}

It would not have been possible without the outstanding creativity of the team of undergraduate students who supported the development of the course in the summer of 2022: Kezziah Ayuno, Joshua Billson, Alinn Martinez, Justin Monteza, and Helen Trinh. We extend our gratitude to the other course instructors (Marianne Morris, Jonathan Schaeffer), the amazing TAs who contributed to the success of this course and to the Department of Computing Science at the University of Alberta for supporting this work. We are thankful to Lorna Simons for proofreading the paper.

\bibliographystyle{ACM-Reference-Format}
\bibliography{references}

\end{document}